\documentclass[10pt, fleqn]{article}
\usepackage{amsmath,amsfonts}
\newcommand{\sumj}{\sum_{j=1}^n\,}
\newcommand{\sumi}{\sum_{i=1}^m\,}
\newcommand{\half}{\mbox{\small{$\frac{1}{2}$}}}
\title{Information and coding discrimination of pseudo-additive entropies (PAE)}
\author{B. H. Lavenda\thanks{email: bernard.lavenda@unicam.it}\\ Universit\'a di Camerino\\ Camerino 62032 (MC) Italy}
\date{{}}
\begin{document}
\maketitle
\begin{abstract}
PAE cannot be made a basis for either a generalized statistical mechanics or a generalized information theory. Either statistical independence must be waived, or the expression of the averaged conditional probability as the difference between the  marginal and joint entropies must be relinquished. The same inequality, relating the PAE  to the R\'enyi entropy, when applied to the mean code length produces an expression that it is without bound as the order of the code length approaches infinity. Since the mean code length associated with the R\'enyi entropy is finite and can be made to come as close to the Hartley entropy as desired in the same limit, the PAE have a more limited range of validity than the R\'enyi entropy which they approximate.
\end{abstract}
\vspace{10pt}
\section{Problems with PAE}
The entropy measure of degree-$\alpha$ \cite{Havrda,Daroczy} and the $\alpha$-norm entropy measure \cite{Arimoto,Lubbe} are PAE insofar as they satisfy a pseudo-additive relation, rather than the additive relation satisfied by the Shannon \cite{Shannon} and R\'enyi \cite{Renyi} entropies. Additivity has been confused with thermodynamic extensivity, and a whole new branch of \lq nonextensive thermodynamics\rq\ has emerged based on the entropy measure of degree-$\alpha$ \cite{Tsallis}. The basis of which is the pseudo-additive relation that expresses the joint entropy in terms of the sums and \emph{products\/} of the marginal entropies. The lack of additivity---even for statistically independent events---has given rise to speculative claims regarding correlations induced by the lack of additivity (extensivity \cite{Abe99}).\par
We shall show that PAE measures of uncertainty are upper or lower bounds on the R\'enyi entropy, depending upon the value of the parameter $\alpha$, and are  incompatible with the rates of transmission of information and channel capacity, expressed in terms of the difference between  marginal  and averaged conditional entropies, or, equivalently,  as the difference between the sum of marginal entropies and the joint entropy. The reasons are:
\begin{enumerate}
\item The averaged conditional entropies are not bounded from above by marginal entropies, which applies in the case of statistically independent events.
\item The conventional rates of transmission of information and channel capacitance can  be negative for values of the characteristic exponent, $\alpha<1$, do not vanish for statistically independent input and output, and not even when the probabilities are equal for correct and incorrect transmission so that the channel does not convey any information at all.
\item  The same approximation that relates the PAE to the R\'enyi entropy when applied to the mean code length associated with the R\'enyi entropy leads to an unbounded mean code in the $\alpha=0$ limit, whereas the former can be made to come as close to the Hartley entropy as desired by the use of extensions of a code.\end{enumerate}
The PAE will be shown to be related to the R\'enyi entropy by the inequality $\ln x\le x-1$, and have no claim in themselves to be considered as genuine expressions of uncertainty for the reasons listed above. Moreover, any viable candidate for an entropy cannot depend upon a parameter, which is different for different systems, since such systems would be incommensurable.\par
\section{PAE in information theory}
The conclusions drawn this section follow from similar results found earlier by Landsberg and Vedral \cite{PTL}, who attempted to insert entropies of degree-$\alpha$ into an information theory framework. Specifically, they found that the channel capacity does not vanish when the input and output are statistically independent, or when the probabilities of correct and incorrect transmission are equal, and it can even become negative, which they thought \lq\lq would be extremely surprising and path-breaking to find a physical system which conveys information in spite of going through a completely destructive channel.\rq\rq\ We shall appreciate that these conclusions are based on the conventional forms of transmission rate and channel capacity that are used for additive entropies. Their expression for the averaged conditional entropy coincides with what Acz\'el and Dar\'oczy \cite{Aczel} have termed the property of \lq  strong additivity of degree-$\alpha$\rq\ [vid. (\ref{eq:add}) below], but, it does not transform into the marginal entropy when the events are statistically independent. A vestige of what was the conditioning variable will be shown to be responsible for all the wrong results.\par
Consider a discrete set of random variables $X$ and $Y$ with
probability distributions $P=(p_1,\ldots,p_m)$ and
$Q=(q_1,\ldots,q_n)$ over sets $(x_1,\ldots,x_m)$, and
$(y_1,\ldots,y_n)$, respectively. Similarly, the two-dimensional
random pair $(X,Y)$ has the joint probability distribution
$\Pi=(\pi_{11},\ldots,\pi_{mn})$, where $\pi_{ij}=\linebreak\Pr\left(X=x_i, Y=y_j\right)$. The conditional
probabilities $p_{ij}$ and $q_{ji}$, are defined by
$\pi_{ij}=p_{ij}q_j=q_{ji}p_i$.\par Papers \cite{Santos,Abe} have
been written proposing a  set of properties that uniquely
characterize the so-called Tsallis entropy
\[
S_1(P)=\frac{1-\sumi p_i^\alpha}{\alpha-1}, \]
in analogy with those that characterize the Shannon entropy.
In addition to the characterization of the entropy of degree-$\alpha$ found in  the original papers \cite{Havrda,Daroczy}, a complete and unique characterization of this  PAE  was given in the 1975
monograph of Acz\'el and Dar\'oczy \cite{Aczel}.\par At the
top of this list is the property of pseudo-additivity of the joint
entropy
\begin{eqnarray}
\lefteqn{S_i^0(x,y)=S_i(x)+\lambda_i(x)S_i(y)}\label{eq:pseudo}\\
&  = & \lambda_i(y)S_i(x)+S_i(y)=S_i(x)+S_i(y)+\tau_iS_i(x)S_i(y), \nonumber
\end{eqnarray}
for statistically \emph{independent\/} events, $x$, and $y$, where
$\lambda_1(x)=\sumi p_i^{\alpha}$, $\lambda_2(y)=\left(\sumj q_j^{\alpha}\right)^{1/\alpha}$, $\tau_1=1-\alpha$, and $\tau_2=(1-\alpha)/\alpha$.  As it stands, the
pseudo-additivity of these entropies, (\ref{eq:pseudo}),
appears as an \emph{ad hoc\/} weighting of one of the marginal entropies by a factor $\lambda_i$. It is only when we consider the
averaging of the conditional entropy that these weights make  some physical sense.\par The averaged conditional entropy is required to be the difference between the joint and marginal entropies:
\begin{equation}
\mathbf{E}\left[S_i(y|X)\right]=S_i(y|x)=S_i(x,y)-S_i(x), \label{eq:E}
\end{equation}
where $\mathbf{E}$ denotes the mathematical expectation. The equality asserts that the uncertainty of $y$, when $x$ is known, is equal to the uncertainty (entropy) of the joint event, $x,y$, less the uncertainty of $x$ \cite{Shannon}. \par
This demands that the averaging  be performed by what has become to be known as averaging with respect to (unnormalized) \lq escort\rq\ probabilities \cite{Beck}. As a result, (\ref{eq:E}) will coincide with the property of  strong additivity of degree-$\alpha$ \cite{Aczel}. The strong additivity of the PAE are
\begin{equation}
S_1(x|y)=\sumj q_j^{\alpha}S_1(x|y_j)=S_1(x,y)-S_1(y),\label{eq:add}
\end{equation}
for the entropy measure of degree-$\alpha$, and  
\begin{equation}
S_2(x|y)=\left(\sumj q_j^{\alpha}\right)^{1/\alpha}S_2(x|y_j)=S_2(x,y)-S_2(y)
 \label{eq:add-bis}
\end{equation}
for the $\alpha$-norm entropy measure.
\par
Any putative expressions for the averaged conditional and joint entropies must satisfy the conditions \cite{Shannon}:
\begin{equation}
S_i(y|x)\le S_i(y) \label{eq:cond-1}
\end{equation}
\noindent
and
\begin{equation}
S_i(x,y)\le S_i(x)+S_i(y). \label{eq:cond-2}
\end{equation}
Expressed in words, (\ref{eq:cond-1}) says that the uncertainty of $y$ is never increased by a knowledge of $x$, and (\ref{eq:cond-2}) affirms that the uncertainty of a joint event can never be greater than the sum of the individual uncertainties.\par
As a consequence of  pseudo-additivity (\ref{eq:pseudo}), conditions (\ref{eq:cond-1}) and (\ref{eq:cond-2}) are replaced by:
\begin{equation}
S_i(y|x)\le\lambda_i(x)S_i(y),\label{eq:cond-1-wrong}
\end{equation}
and
\begin{equation}
S_i(x,y)\le S_i(x)+S_i(y)\le S_i^0(x,y)-\tau_iS_i(x)S_i(y),\label{eq:cond-2-wrong}
\end{equation}
respectively, since the joint entropy is greatest when the events are statistically independent. According to (\ref{eq:cond-1-wrong}), it is not true for $\alpha<1$ ($\lambda_i>1$) that the averaged conditional entropy is inferior, or, at most equal, to the marginal entropy. When $x$ does not condition $y$, the right-hand side of (\ref{eq:cond-1-wrong}) should be independent of $x$, which it is not. A vestige of the conditioning variable, even when it is no longer conditioning, demands a physical explanation. The joint entropy must be greatest in the case of statistically independent events, and (\ref{eq:cond-2-wrong}) shows that for $\alpha>1$ this may not necessarily be so. Since Landsberg and Vedral based their analysis on the validity of the expression for the averaged conditional entropy, (\ref{eq:add}), it is no wonder that they came to absurd conclusions regarding the transmission rate and channel capacity when the entropy measure of degree-$\alpha$ is used instead of  additive entropies. \par
The expression for the averaged conditional entropy in (\ref{eq:add-bis}) was rejected by Boekee and Lubbe \cite{Lubbe} precisely because it does not satisfy the equality in (\ref{eq:cond-1}) in the case of statistically independent events. A classical extension of Minkowski's inequality \cite{BB},
\[\left(\sumj q_j^{\alpha}\right)^{1/\alpha}\le
\sumi p_i\left(\sumj q_{ji}^{\alpha}\right)^{1/\alpha},\]
for $\alpha>1$, was used to show that  ordinary averaging, using weights, $p_i$, of the conditional $\alpha$-norm measure of uncertainty satisfies condition (\ref{eq:cond-1}). The price to be paid is that (\ref{eq:E}) no longer holds.\par
The expressions for the rate of transmission of information and channel capacity that should have been used are:
\begin{equation}
R_i=S_i(x)-\frac{S_i(x|y)}{\lambda_i(y)},\label{eq:R}
\end{equation}
and 
\begin{equation}
C_i=\max_x\left\{S_i(x)-\frac{S_i(x|y)}{\lambda_i(y)}\right\},\label{eq:C}
\end{equation}
respectively, where the maximum is with respect to all possible information sources used as input to the channel. The division of the averaged conditional entropies by the weighting factors $\lambda_i(y)$ together with the property of strong additivity, (\ref{eq:add}) and (\ref{eq:add-bis}), insure that when the
inputs and outputs are statistically independent, or when there is equal probability for correct or incorrect transmission, the transmission rate and the channel capacity vanish.\par However, this does not come without a cost: What is compromised is the relation between the marginal and joint probabilities in that the transmission rate is proportional to the physically uninterpretable difference $\lambda_i(y)S_i(x)+S_i(y)-S_i(x,y)$, rather than the difference between the sum of marginal entropies and the joint entropy, $S_i(x)+S_i(y)-S_i(x,y)$, as it would be were the entropies additive \cite[p. 38]{Shannon}. Although the transmission rate, (\ref{eq:R}), and hence the channel capacity, (\ref{eq:C}), appear to be increased (decreased) beyond their conventional values for $\alpha<1$ ($\alpha>1$), they can never become negative in the range $\alpha<1$, as they would be---above all for statistically independent input and output---were the conventional forms of the $R$ and $C$ for additive entropies to be employed \cite{PTL}.
\par
The conclusion that we are forced into accepting is that the entropy measure of degree-$\alpha$, as well as the $\alpha$-norm entropy measure, cannot be used as valid expressions for a true entropy. We will now show that they are related to an interpolation formula which varies between the Shannon entropy when $\alpha=1$ to the Hartley entropy when $\alpha=0$. It is quite remarkable that the latter does not require equal probabilities, as the PAE would---in the $\alpha=1$ limit rather than the $\alpha=0$ limit.\par The interpolation formula is the R\'enyi measure of uncertainty:
\[
S_R(x)=\frac{1}{\tau_2}\ln\left(\sumi p_i^{\alpha}\right)^{1/\alpha},
\]
which we have written in a form that displays the intimate relation between the entropy measure of degree-$\alpha$ and the $\alpha$-norm measure. On the strength of inequality $\ln x\le x-1$, it is readily seen that the R\'enyi entropy can never be greater than the PAE
for $\alpha<1$, and never smaller than the PAE  for $\alpha>1$ \cite{Arimoto}.
\par
The averaged conditional R\'enyi entropy,
 \begin{equation}
S_R(y|x)=S_R(x,y)-S_R(x)=\frac{1}{\tau_2}\ln\left(\sumi\sumj q_{ji}^{\alpha}\frac{
p_i^{\alpha}}{\sumi p_i^{\alpha}}\right)^{1/\alpha} \label{eq:Renyi-con},
\end{equation}
looks like a normalized escort average of the conditional probabilities to the power $\alpha$. Applying the inequality $\ln x\le x-1$  to the right-hand side of (\ref{eq:Renyi-con}) we get
\begin{eqnarray*}
\lefteqn{S_R(y|x)\le}\\
& & \frac{1}{\tau_2}
\frac{\left(\sumj\sumi\pi_{ij}^{\alpha}\right)^{1/\alpha}
-\left(\sumi p_i^{\alpha}\right)^{1/\alpha}}
{\left(\sumi p_i^{\alpha}\right)^{1/\alpha}}=
\frac{S_2(x,y)-S_2(x)}{\lambda_2(x)}
\end{eqnarray*}for $\alpha<1$, and the reverse inequality for $\alpha>1$.
Alternatively, this can be written in terms of the difference of the joint and marginal entropies of degree-$\alpha<1$,
\begin{eqnarray}
\lefteqn{S_R(y|x)\le}\nonumber\\
& & \frac{1}{\tau_1}\frac{\sumj\sumi\pi_{ij}^{\alpha}-\sumi p_i^{\alpha}}{\sumi p_i^{\alpha}}=\frac{S_1(x,y)-S_1(x)}{\lambda_1(x)}. \label{eq:Abe}
\end{eqnarray}
\par
Abe \cite{Abe} called the ratio on the right-hand side of (\ref{eq:Abe}) the averaged conditional entropy of degree-$\alpha$, instead of the numerator, and obtained it using a normalized escort averaging of the conditional entropy of degree-$\alpha$. Notwithstanding the fact that this averaged conditional entropy is no longer the difference between the joint and marginal entropies, conditions (\ref{eq:cond-1}) and (\ref{eq:cond-2}) are satisfied, and amount to considering expressions (\ref{eq:R}) and (\ref{eq:C}) for the rate of transmission of information and the channel capacity. The normalizing factor eliminates the vestigial weighting factor in (\ref{eq:cond-1-wrong}), that depends what was formerly the conditioning variable. Consequently, the channel capacity, as defined by (\ref{eq:C}), will always be  positive semidefinite,  vanishing  either when the input and output are statistically independent, or when  the probabilities for correct and incorrect channel transmission are equal.\par To illustrate the latter, it suffices to consider a  binary  symmetric channel. Let $P$ be the probability of incorrect transmission, i.e., $q_{11}=q_{22}=1-P$ and $q_{12}=q_{21}=P$. If $p$ ($q$) is the probability of the input (output) $x=0$ ($y=0)$ and $1-p$ ($1-q$) that of $x=1$ ($y=1$), then the transmission rate, (\ref{eq:R}), is
\[R=\frac{p^{\alpha}+(1-p)^{\alpha}}
{q^{\alpha}+(1-q)^{\alpha}}\cdot\frac{q^{\alpha}+(1-q)^{\alpha}-(1-P)^{\alpha}-P^{\alpha}}{1-\alpha},\]
and the channel capacity,  
\[C=\frac{2^{1-\alpha}-(1-P)^{\alpha}-P^{\alpha}}{1-\alpha},\]
vanishes for $P=\half$, because the channel does not convey any information.\par
Therefore, at the expense of destroying the relation between the marginal and joint entropies, or that between the averaged conditional entropy and the difference between the joint and marginal entropies, the \lq unphysical\rq\ results found in \cite{PTL} are avoided. But this makes it only  more apparent that the PAE are not fundamental quantities upon which a theory of information, or a generalized statistical mechanics, can be constructed.\par
\section{Characterization of PAE through a coding problem}
In order to show the limitations of the entropy measure of degree-$\alpha$ and the $\alpha$-norm measure, consider possible measures of the length of a code. We have an alphabet of $D$ symbols, $d_1,\ldots,d_D$ into which the input symbols are to be encoded. Let $p_1,\ldots,p_m$ be the probabilities of $m$ input symbols, $x_1,\ldots,x_m$ from an information source. To each $x$ we wish to associate a sequence of the $d$'s, with the only restriction that no sequence of the $d$'s shall be obtainable from a shorter sequence by the addition of more terms to the shorter \cite{Feinstein}. The length $\ell_i$ of a sequence that is to be associated, in some way, with $x_i$ will have an average length $\sumi p_i\ell_i$.\par According to Kraft's theorem, there will be a uniquely decipherable code if and only if \cite{Feinstein}
\begin{equation}
\sumi D^{-\ell_i}\le1\label{eq:Kraft}
\end{equation}
is satisfied. The condition for equality is $\ell_i=-\ln_Dp_i$, since the $p_i$ form a complete set. But, this is not the only optimal  code length.
\par
There may be many different codes whose lengths satisfy the Kraft inequality, (\ref{eq:Kraft}). In order to determine the optimum code it is necessary to consider the mean code length, and to minimize it.  Such a procedure is valid when the  \lq cost\rq\ of using a sequence of length $\ell_i$ is proportional to the mean length of a code, $\sumi p_i\ell_i$. However, a more general cost would show an exponential dependence on the lengths, $\ell_i$ \cite{Campbell}
\begin{equation}
C(\tau_2)=\sumi p_iD^{\tau_2\ell_i},\label{eq:Campbell}
\end{equation}
and span a range of optimal code lengths as a function of the parameter $\tau_2$.\par If $\tau_2$ is restricted to the range $0\le\tau_2\le\infty$ ($0\le\alpha\le1$), the length of a code \cite{Campbell}
\begin{equation}
L(\tau_2)=\frac{1}{\tau_2}\ln_DC(\tau_2),\label{eq:L}
\end{equation}
can vary from
\begin{equation}
L(0)=\sumi p_i\ell_i \label{eq:average}
\end{equation}
as $\tau_2\rightarrow0$ ($\alpha\rightarrow1$) to 
\begin{equation}
L(\infty)=\lim_{\tau_{2}\rightarrow\infty}L(\tau_2)=
\max_{1\le i\le m}\ell_i\label{eq:ell}
\end{equation}
as $\tau_2\rightarrow\infty$ ($\alpha\rightarrow0$), 
since for large $\tau_2$,
\begin{equation}
\sumi p_iD^{\tau_{2}\ell_{i}}\doteq p_kD^{\tau_{2}\ell_{k}},\label{eq:ell-ext}
\end{equation}
where $\ell_k$ is the largest of the numbers $\ell_1,\ldots,\ell_m$, and $p_k$ is its probability. \par In the former case, $\ell_i\ge-\ln_Dp_i$, and the Shannon entropy becomes the lower limit to the mean code length \cite[p. 30]{Shannon}
\[
L(0)=\sumi p_i\ell_i\ge-\sumi p_i\ln_Dp_i.\]
In the latter case, H\"older's inequality can be used to obtain
\cite{Campbell}
\begin{equation}\sumi p_iD^{\tau_2\ell_i}\ge\left(\sumi p_i^{\alpha}\right)^{1/\alpha},\label{eq:D}
\end{equation}
so that
\begin{equation}
L(\tau_2)\ge\frac{1}{\tau_2}\ln_D\left(\sumi p_i^{\alpha}\right)
^{1/\alpha},\label{eq:Renyi-bound}
\end{equation}
for $0<\tau_2<\infty$. And over this range, the R\'enyi measure of uncertainty becomes the lower limit for the mean code length.\par Moreover, the optimal code lengths can be expressed in terms of the normalized escort probabilities as
\[D^{-\ell_i}=\frac{p_i^{\alpha}}{\sumi p_i^{\alpha}},\]
which again gives the equality in (\ref{eq:Kraft}). Since $\alpha$ lies in the interval $(0,1)$, $\ell_i>-\alpha\ln_Dp_i$, and no conclusion may be reached in comparison with the lengths $\ell_i=-\ln_Dp_i$, when the mean length is given by the weighted average.\par
Applying inequality $\ln x\le x-1$ to the definition of the mean length (\ref{eq:L}) gives a new mean code length which is directly proportional to the cost, viz.,
\begin{equation}
L(\tau_2)=\frac{C(\tau_2)-1}{\tau_2}\ge\frac{1}{\tau_2}\left\{\left(\sumi p_i^{\alpha}\right)^{1/\alpha}-1\right\},
\label{eq:L-bis}
\end{equation} where the inequality follows from (\ref{eq:D}). Thus, the $\alpha$-norm entropy measure is the lower limit to the mean code length of order $\tau_2$, which is proportional to the cost (\ref{eq:Campbell}), instead of its logarithm as in (\ref{eq:L}).\par The reason for calling (\ref{eq:L-bis}) a length is that if all the code lengths were equal to $\ell$,
\[L(\tau_2)=\frac{D^{\tau_2\ell}-1}{\tau_2},\]
and in the limit as $\tau_2\rightarrow0$, it would reduce simply to $L(0)=\ell$. More generally, for unequal code lengths, (\ref{eq:L-bis}) becomes the weighted average (\ref{eq:average}) in the same limit, and the $\alpha$-norm entropy measure becomes the Shannon entropy.\par However, in the limit as $\tau_2\rightarrow\infty$, instead of (\ref{eq:ell}), whose finiteness is due the presence of the logarithm in the numerator, we now get \begin{equation}
L(\tau_2)=\frac{p_kD^{\tau_2\ell_{k}}}{\tau_2}, \label{eq:infty}
\end{equation}
on account of (\ref{eq:ell-ext}). There is no finite upper limit to (\ref{eq:infty}).\par The lack of an upper bound on the mean code length in the $\tau_2=\infty$ limit is in contradiction with the well-known bounds on the average code length for $M$-extensions of a code, which is all possible concatenations of the $m$ symbols of the original source code. In the limit $\alpha\rightarrow0$, the R\'enyi measure of uncertainty becomes the Hartley entropy, and,  for the $M$th extension,  the mean code length is bounded by \cite{Campbell}:
\begin{equation}
S_0\le\frac{L_M(\infty)}{M}\le S_0+\frac{1}{M}.\label{eq:H}
\end{equation}
The total entropy is $M$ times as large as the original Hartley entropy, $S_0=\ln_Dm$,  and the averaged code length  $L_M(\infty)=\max\ell(s)$, where $s$ is an input sequence of length $M$. By choosing $M$ sufficiently large, the average code length per extension, $L_M(\infty)/M$, can be made to come as close to the Hartley entropy as desired in the $\tau_2=\infty$ limit.\par 
Consequently, the approximation of R\'enyi measure of uncertainty by the PAE is not valid in the $\tau_2=\infty$ ($\alpha=0$) limit. Whereas the R\'enyi entropy transforms into the Hartley entropy, and provides upper and lower bounds on the mean code length per extension, as given by (\ref{eq:H}), the PAE do not provide similar bounds on the code length (\ref{eq:infty}) because the latter is unbounded in the $\tau_2=\infty$ limit. As a matter of fact, the PAE reduce to the Hartley entropy in the $\alpha=1$ limit, but, only after the probabilities have been set equal. Therefore the PAE have a more  limited range of validity than the R\'enyi entropy which they approximate. This together with their problems of handling  statistical independent events and correlations among statistical dependent ones do not make them suitable as a basis for either a generalized information theory, or a generalized statistical mechanics.

\end{document}